\documentclass[twocolumn,amsmath,amssymb,showpacs]{revtex4}

\usepackage{graphicx}

\begin{document}

\title{Vortex pinning by the point potential in topological superconductors: a scheme for braiding Majorana bound states}

\author{Hai-Dan Wu and Tao Zhou}
\email{tzhou@nuaa.edu.cn}
\affiliation{College of Science, Nanjing University of Aeronautics and Astronautics, Nanjing 210016, China.}

\date{\today}
\begin{abstract}
We propose theoretically an effective scheme for braiding Majorana bound states by manipulating the point potential.
The vortex pinning effect
is carefully elucidated. This effect may be used to control the vortices and Majorana bound states in topological superconductors.
The exchange of two vortices induced by moving the potentials is simulated numerically.
The zero-energy state in the vortex core is robust with respect to the strength of the potential. The Majorana bound states in a pinned vortex are identified numerically.
\end{abstract}
\maketitle

\section{introduction}

Topological superconductors have been studied intensively because of their exotic properties~\cite{masa1}. They are characterized by a bulk superconducting gap and topologically
protected gapless states at the system edges, which are subject to the excitation of Majorana bound states (MBSs). In the past, the realization of MBSs in various topological superconducting systems has attracted broad interest~\cite{read,lfu,lutc,oreg,sau,alic,wil,rok,den,das,mou,jhlee,sun,mtdeng,nadj,qlhe}. Theoretically, a straightforward mechanism is that MBSs appear at the system edges and vortex cores of a $p+ip$ superconductor~\cite{read}. However, searching for a superconducting material with $p$-wave pairing symmetry is a great challenge. Therefore, many more realistic models have also been proposed~\cite{lfu,lutc,oreg,sau,alic}. One well-known proposal is that a $p+ip$ superconductor is equivalent to a system with an $s$-wave superconductor coupled to a topological insulator. In the presence of a magnetic field, MBSs will be excited in the vortex cores~\cite{lfu}. Another promising proposal is that MBSs can be realized in a sandwich system that includes $s$-wave pairing,  spin-orbital interaction and a Zeeman field. Then, by tuning the Fermi level, this system can also be made equivalent to a $p+ip$ superconductor~\cite{lutc,oreg,sau,alic}. The two proposals introduced above have both been realized experimentally, and possible signatures of MBSs have been reported based on the zero-bias peak in the differential tunneling conductance measured in scanning tunneling microscopy experiments~\cite{wil,rok,den,das,mou,jhlee,sun,mtdeng}.

One of the most important features of MBSs is that they usually obey non-Abelian statistics~\cite{iva}. This feature is of fundamental interest, and a direct experimental demonstration of non-Abelian statistics may provide definite evidence of the existence of MBSs. Moreover, this property has potential applications in topological quantum computations.
However, it is rather difficult to find an experimental platform with which to realize the braiding of MBSs and demonstrate
their non-Abelian statistics. Although there have been many theoretical proposals~\cite{jali,jdsau,cla,hal,kli,skkim,jli,aas,amat}, to date, this remains an open question.
Very recently, it was reported that the exchange of Majorana zero modes can be simulated using a photonic system~\cite{jsxu,jxu}.
However, there has still been no experimental realization of the manipulation MBSs to demonstrate their non-Abelian statistics in topological superconductors.

 \begin{figure}
  \centering
  \includegraphics[width=0.45\textwidth]{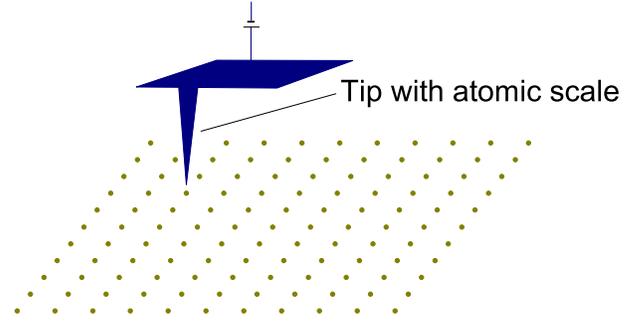}\\
  \caption{(Color online) Schematic diagram of a method for generating a controllable point potential.}\label{fig1}
\end{figure}

The interaction between a vortex and a point potential has previously attracted broad interest~\cite{han,zhu,kim,zhou2}.
It has been demonstrated numerically that the center of a vortex may be attracted and pinned by a point potential. Such a pinning effect is of both theoretical and practical interest.
Another important issue is the interplay between the vortex bound states and the potentials. However, numerically different results have been reported in different families of superconductors~\cite{han,zhu,kim,zhou2}.
In topological superconductors, the pinning effect induced by a local potential and its possible applications have not previously been explored.
In fact, this possible pinning effect may offer a potential method of controlling vortices.
In particular, it has been verified theoretically that
the zero energy bound states are topologically protected and that they are not affected by local potentials~\cite{volo,tanu,masa,masa2,smit,guo}.
If the zero-energy bound states are subjected to the MBS excitation, then the pinning effect may provide an effective means of manipulating the MBSs in topological superconductors. The braiding of the MBSs may be realized by manipulating the potentials.

In this paper, motivated by the above considerations, we
study theoretically the effect of a point potential on a vortex state using the self-consistent Bogoliubov-de
Gennes (BdG) technique. An effective controllable point potential can be produced by means of atomic force microscopy~\cite{gie}. As shown in Fig.~1, we consider an atomic-scale tip mounted on a movable device. The potential originates from the interaction between atoms at the front of the tip and the quasiparticles in the sample. We propose that the Coulomb interaction may produce a suitable
potential. The strength of the potential can be well controlled, and due to the screening effect, the Coulomb potential is rather localized and may be treated as a point potential.
Starting from an effective model describing topological superconductors~\cite{lutc,oreg,sau,alic},
the vortex pinning effect and the critical pinning distance are investigated, to seek an effective method of controlling and manipulating the MBSs. 
The exchange of two vortices induced by slowly moving two point potentials is simulated. We also check numerically that the zero mode is robust against the potentials.
Meanwhile, it is clearly verified that the zero mode is indeed subject to two separate MBSs.

The rest of the paper is organized as follows.
In Sec. II, we introduce
the model and derive the formalism. In Sec. III, we
report numerical calculations and discuss the obtained
results. Finally, we present a brief summary in Sec. IV.

\section{Model and Hamiltonian}
  Following Refs.~\cite{lutc,oreg,sau,alic}, an effective theoretical model that describes a topological superconductor includes a hopping term, spin-orbital coupling, a Zeeman field, and an $s$-wave superconducting pairing term.
 Considering the terms listed above, our starting model is expressed as
\begin{equation}
H=H_t+H_{SO}+H_{SC}.
\end{equation}
Here, $H_t$ includes the hopping term, the chemical potential term, the Zeeman field, and an additional point potential term; it is expressed as,
\begin{eqnarray}
H_t=&-\sum_{\langle {\bf ij}\rangle} t_0\Phi_{\bf ij} (c^{\dagger}_{{\bf i}\sigma} c_{{\bf j}\sigma}+h.c.)+\sum_{{\bf i}\sigma}(\sigma h-\mu)c^{\dagger}_{{\bf i}\sigma} c_{{\bf i}\sigma}\nonumber\\&+\sum_\sigma V_i c^{\dagger}_{{\bf i_0}\sigma} c_{{\bf i_0}\sigma}  .
\end{eqnarray}
$\langle {\bf ij}\rangle$ represents the nearest-neighbor sites. $t_0$ is the nearest-neighbor hopping constant.
In the presence of a magnetic field, we have $\Phi_{\bf ij}=\exp [i\pi/\phi_0]\int ^{\bf R_j}_{\bf R_i} A({\bf r})d{\bf r}$, where $\phi_0$ is the superconducting flux quantum and
$A = (-By,0,0)$ is the vector potential in the Landau
gauge.
$h$ is the Zeeman field. The sign $\sigma$ is ``$+$" for the spin-up state and ``$-$" for the spin-down state. $\mu$ is the chemical potential.
We consider a point potential at site ${\bf i_0}$ with a strength of
$V_i$.

$H_{SO}$ is the spin-orbital interaction, expressed as
\begin{eqnarray}
H_{SO}=&\sum_{\bf i}(i\lambda \Phi_{\bf ij} c^{\dagger}_{{\bf i}\uparrow}c_{{\bf i}+\hat{x}\downarrow}+i\lambda \Phi_{\bf ij} c^{\dagger}_{{\bf i}\downarrow}c_{{\bf i}+\hat{x}\uparrow}+h.c.\nonumber\\
&+\lambda \Phi_{\bf ij} c^{\dagger}_{{\bf i}\uparrow}c_{{\bf i}+\hat{y}\downarrow}-\lambda \Phi_{\bf ij} c^{\dagger}_{{\bf i}\downarrow}c_{{\bf i}+\hat{y}\uparrow}+h.c.),
\end{eqnarray}
where $\lambda$ is the spin-orbital coupling strength.

$H_{SC}$ is the superconducting pairing term, expressed as
\begin{equation}
H_{SC}=\sum_{\bf i}(\Delta_{\bf ii}c^{\dagger}_{{\bf i}\uparrow}c^{\dagger}_{{\bf i}\downarrow}+h.c.).
\end{equation}
Here, $\Delta_{\bf ii}$ is the on-site pairing order parameter, that originates from an on-site attractive interaction.

The above Hamiltonian can be diagonalized by solving the BdG equations self-consistently:
\begin{eqnarray}
\sum_{\bf j}\left(
\begin{array}{cccc}
H_{{\bf ij}\uparrow\uparrow}&H_{{\bf ij}\uparrow\downarrow}&\Delta_{\bf jj}&0\\
H_{{\bf ij}\downarrow\uparrow}&H_{{\bf ij}\downarrow\downarrow}&0&-\Delta_{\bf jj}\\
\Delta^{*}_{\bf jj}&0&-H_{{\bf ij}\downarrow\downarrow}&-H^{*}_{{\bf ij}\downarrow\uparrow}\\
0&-\Delta^{*}_{\bf jj}&-H^{*}_{{\bf ij}\uparrow\downarrow}&-H_{{\bf ij}\uparrow\uparrow}
\end{array}
\right)
\Psi_{\bf j}^{\eta}
=E_{\eta}\Psi_{\bf j}^{\eta},
\end{eqnarray}
where $\Psi_{\bf j}^{\eta}=(u_{\bf j\uparrow}^{\eta},u_{\bf j\downarrow}^{\eta},v_{\bf j\downarrow}^{\eta},v_{\bf j\uparrow}^{\eta})^T$.
$H_{{\bf ij}\sigma\sigma}$ and $H_{{\bf ij}\sigma\bar{\sigma}}$ $(\sigma\neq\bar{\sigma})$ are obtained from $H_t$ and $H_{SO}$, respectively.

The order parameters $\Delta_{\bf jj}$ are calculated as follows:
\begin{equation}
\Delta_{\bf jj}=\frac{V}{2}\sum_{\eta}u^{\eta}_{{\bf j}\uparrow}v^{\eta\ast}_{{\bf j}\downarrow}\tanh(\frac{E_{\eta}}{2K_{B}T}),
\end{equation}
where $V$ is the pairing strength.

The BdG equations are solved self-consistently as follows.
First, we input a set of initial order parameters $\Delta_{\bf ii}$ and diagonalize the Hamiltonian matrix.
Second, we use the obtained eigenfunctions
and eigenvalues to calculate the new order parameters as expressed in Eq.(6). This procedure is repeated with
the updated order parameters until the convergence criterion is satisfied.

In the results reported below, we use the lattice constant $a$ and the nearest-neighbor hopping constant $t_0$ as the units of length and energy, respectively. Without loss of generality, the other parameters are set to $\mu=-4$, $h=0.6$, and $V=5$, respectively. Then, the Hamiltonian given in Eq.(1) is an effective model that describes the corresponding topological superconductor.
The superconducting gap is calculated self-consistently on the $N=48\times48$ lattice with periodic boundary conditions. The magnetic field $B$ is chosen to be $B={2\phi_0}/(N a^2)$. 
Under these conditions, two vortices should exist in the system.
We have checked numerically that our main results are not qualitatively different for different input parameters.

 \begin{figure}
  \centering
  \includegraphics[width=0.45\textwidth]{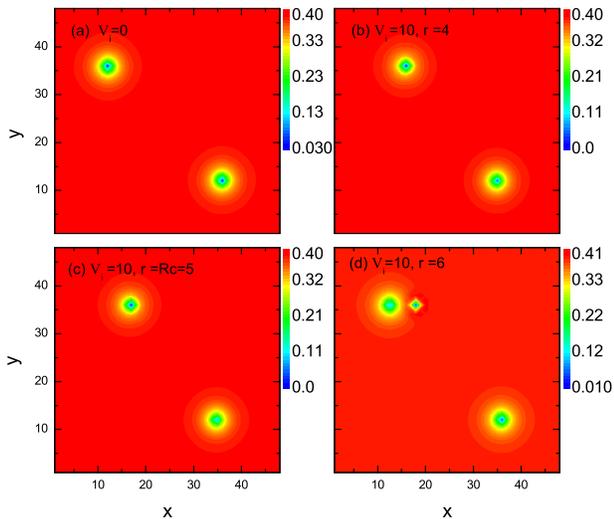}\\
  \caption{(Color online) (a) The intensity plot of the superconducting gap without the point potential. Two vortices are located at the sites $(12,36)$ and $(36,12)$. (b-d) Spatial variations of the superconducting gap with the existence of a single point potential at the site $(16,36)$, $(17,36)$, and $(18,36)$, respectively. }\label{fig1}
\end{figure}

\section{Results and Discussions}

We first investigate the pinning effect and the pinning distance. The spatial
variations of the superconducting pairing order parameter without an additional point potential and with a single potential at different sites are displayed in Fig.~2.
For the case without the point potential, we chose random initial values as the input order parameters.
The intensity plot of the superconducting order parameter without the point potential is presented in Fig.~2(a).
For the cases with one point potential in the system, we used the convergent self-consistent results for $V_i=0$ as the input parameters.
The corresponding numerical results are presented in Figs.~2(b-d).

 Without the potential, as seen in Fig.~2(a), two vortices appear,
centered at the sites (12,36) and (36,12).
In the presence of a single point potential near the core of a vortex,
 as presented in Figs.~2(b) and 2(c), the vortex center moves and is pinned.  A critical
distance $R_c$ between the potential site and the previous center position of the core
 can be defined. When the distance is smaller than $R_c$, the
center of the vortex core will be dragged to the potential site. For
the present set of parameters, $R_c$ is approximately five times the lattice constants.
 When the distance $r$ is larger than $R_c$, as shown in Fig.~2(d), the vortex and the potential will be
separated, with the vortex center remaining at its previous site, $(12,36)$.

\begin{figure}
  \centering
  \includegraphics[width=0.45\textwidth]{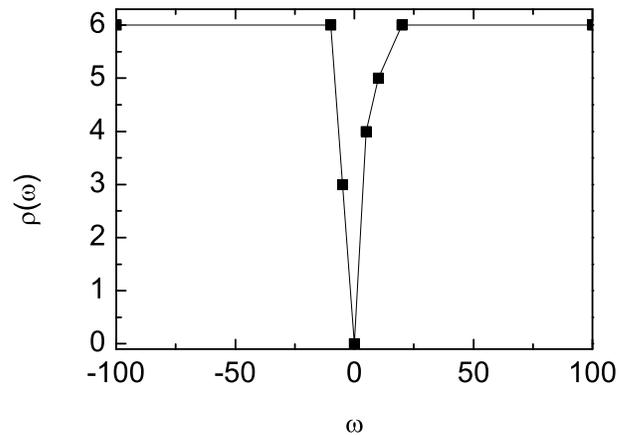}\\
  \caption{The critical pinning distance $R_c$ as a function of the potential strength $V_i$.}\label{fig3}
\end{figure}

Let us present a more quantitative study of the pinning effect. Generally, the critical pinning distance $R_c$
is insensitive to the direction of the potential site with respect to the vortex center. It is mainly determined by the potential strength $V_i$.
The critical distance $R_c$ is plotted as a function of the potential $V_i$ in Fig.~3, where $V_i<0$ for an attractive potential and $V_i>0$ for a repulsive one.
As shown, $R_c$ increases as $V_i$ increases and saturates at approximately six times the lattice constant for strong potentials.

In our present model, a critical Zeeman field $h_c$ exists, where $h_c=0.48$. The system transits into the topologically trivial phase when the Zeeman field is smaller than $h_c$.
We have checked numerically that the critical pinning distance $R_c$ does not change as $h$ crosses $h_c$. Thus, the vortex pinning effect described above is not a unique property of topological superconductors. Indeed, similar effects have been previously reported
 in other families of superconductors~\cite{han,zhu,kim,zhou2}.
 However, we emphasize here that for topological superconductors, the vortex pinning effect may be particularly interesting and practical useful.
 As presented in Fig.~1, when the potential site is far away from the center of a vortex, it does not affect that vortex. However, if the point potential
 can be well controlled and is moved slowly away from the vortex center, we expect that the vortex will be dragged along with the potential.
 If the vortex density is sufficiently low that the interaction between vortices is negligibly small,
 then, in principle, the vortex may be dragged to any site in the system.
Because the vortices in topological superconductors are usually associated with MBS excitation, this dragging effect may be used to
achieve the braiding of MBSs.

We attempt to manipulate the vortices by moving the potentials in accordance with the above assumption. As the first step, we assume that a single potential $(V_i=10)$ exists at the center of a vortex ${\bf r}={\bf R_0}$ [${\bf R_0}=(12,36)$].
The site-dependent order parameters can be obtained by solving the BdG equations. We then move the potential from ${\bf R_0}$ to its neighbor site ${\bf R_1}$. In this step, we use the self-consistent results
for ${\bf r}={\bf R_0}$ as the initial input parameters to calculate the updated order parameters for ${\bf r}={\bf R_1}$. Then, the potential is moved to ${\bf R_2}$, and the self-consistent results
for ${\bf r}={\bf R_1}$ are used as the new input parameters. In this way, we move the vortex from the site ${\bf R_0}=(12,36)$ to the site ${\bf R_f}=(28,36)$ by slowly changing the point potential.
The intensity plot of the order parameters after the potential has been moved to the site $R_f$ is presented in Fig.~4(a).
Note that here, the distance between ${\bf R_f}$ and ${\bf R_0}$ reaches sixteen times lattice constants, much larger than the critical pinning distance.
 Our results indicate that the vortex can indeed be well controlled if we move the potential sufficiently slowly.

We now simulate the exchange of the two vortices, which is achieved by manipulating two point potentials.
As displayed in Fig.~4(b), initially, two pinned vortices exist
at the sites $(12,36)$ (vortex A) and $(36,12)$ (vortex B). Then, the potentials are moved slowly along the directions of the corresponding arrows. As the positions of the potentials move, the previous self-consistent results are taken as the new initial input parameters. Several intermediate results are displayed in Figs.~4(c-e). The final result is presented in Fig.~4(f), where the vortices $A$ and $B$ have moved to the sites $(36,12)$ and $(12,36)$, respectively. Therefore, using this method, we can successfully exchange the two vortices.

We have demonstrated that vortices can be well controlled by means of the manipulation of point potentials.
However, one essential question is whether the MBSs will survive for pinned vortices. According to previous theoretical calculations, the zero-energy bound states should be robust with respect to a point potential. However, the zero-energy bound states are not necessarily subject to the MBS excitation. Therefore, it is still important to check numerically whether the MBSs are robust when a vortex is pinned by a point potential. We consider one point potential located at the site (12,36) $(V_i=10)$. The eigenvalues obtained by diagonalizing the Hamiltonian are plotted in Fig.~5(a). Two zero-energy
eigenvalues are clearly seen from the numerical results, protected by a minigap of approximately 0.1. The existence of topologically protected zero-energy states usually indicates MBS excitation. We will discuss this issue further in relation to the results presented below.

\begin{figure}
  \centering
  \includegraphics[width=0.45\textwidth]{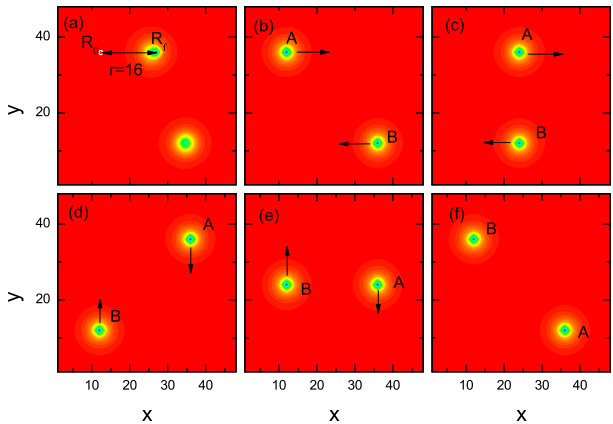}\\
  \caption{The intensity plots of the superconducting gap as an illustration of the manipulation of the vortices. (a)
  Simulation of dragging one vortex from $(12,36)$ to $(28,36)$ by slowly moving a point potential.
  The spatial distribution of the superconducting gap with the point potential at the site $(28,36)$ is displayed.
(b-f) Simulation of the exchange of the two vortices induced by manipulating two point potentials.     }\label{fig4}
\end{figure}

For the Hamiltonian with the superconducting pairing term, the eigenvalues $E$ and $-E$ always appear in pairs due to particle-hole symmetry, where the corresponding eigenvectors are denoted by $C$ and $C^\dagger$, respectively.
 For a zero energy particle $(E=0)$, $C$ and $C^\dagger$ are obviously degenerate. Therefore, their superpositions are also eigenvectors of the Hamiltonian.
 Then, one can obtain two Majorana particles, $\gamma_{1,2}$, from the particle operators $C$ and $C^\dagger$, expressed as $\gamma_1=(C+C^\dagger)/\sqrt{2}$ and $\gamma_2=i(C^{\dagger}-C)/\sqrt{2}$. Here, $C$ and $C^{\dagger}$ are obtained numerically from the eigenvectors corresponding to the zero-energy eigenvalues. Thus, the existence of MBSs can be studied and identified numerically. The spatial distributions of the two MBSs are plotted in Figs.~5(b) and 5(c).
  As is seen, two well-separated MBSs are identified. Each is bounded by its corresponding vortex.
  For a pinned vortex [Fig.~5(b)], the spectral function of the MBS is suppressed significantly at the core center. The maximum spectral weight appears at nearest-neighbor or next-nearest-neighbor sites to the potential. For an unpinned vortex, as seen in Figs.5(c), the MBS ($\gamma_2$) is localized at the vortex core, and the spectral weight decays rapidly at distances farther from the core center.

\begin{figure}
  \centering
  \includegraphics[width=0.45\textwidth]{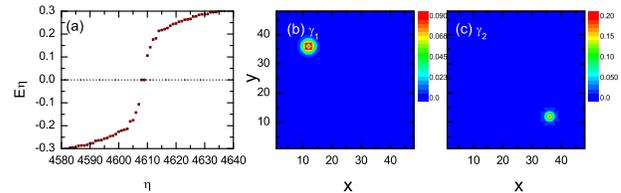}\\
  \caption{(a) The eigenvalues ($E_\eta$) of the Hamiltonian matrix when one vortex is pinned by a point potential with $V_i=10$.
  The corresponding spatial distributions of the two separate MBSs are displayed in panels (b) and (c).
 }\label{fig5}
\end{figure}

Finally, we would like to discuss the practical feasibility of our scheme. First, it is important to clarify whether the results are qualitatively the same
when a local potential of a finite size is considered.
We have checked numerically that the zero-energy states and the MBSs are robust to local potentials of larger scale. Moreover, the critical pinning distance $R_c$ increases significantly when offsite potentials are added. Therefore, in fact, the MBSs are easier to control in a real system, in which the actual potential may cover several sites. Second, our scheme requires the vortex density to be low enough that the dragging force will dominate over the interaction between vortices and that other vortices will not be affected when one vortex is dragged by a potential.
The vortex density is determined by the magnetic field strength $B$ and can be calculated theoretically. Under the assumption that 
the lattice constant $a$ is about $3${\AA}, the magnetic field $B$ is estimated to be approximately $20$T in our present work. This is much larger than the experimental value. In a real system, the vortex lattice can survive for a rather low magnetic field ($0.02$T)~\cite{hes}. With this field strength the distance between two neighboring vortices is rather
large (approximately $3460${\AA}). Thus, we can safely conclude that in a real system, the vortex density can be sufficiently low to allow the MBSs to
be well controlled by potentials.
Therefore, we expect that our present scheme is reasonable and is possible to realize in a real system.

\section{summary}
In summary, we have studied theoretically the effect of a point potential on the vortex states in topological superconductors.
The center of a vortex is dragged to the potential when the potential is close to the vortex. This pinning effect may be used to control vortices. We simulated the exchange of two vortices induced by slowly moving two
point potentials.
The zero-energy states and the MBS excitation for a pinned vortex were identified numerically.
Therefore, we have presented an effective scheme for the braiding of MBSs.

We thank Wei Chen for helpful discussions.
This work was supported
by the NSFC under the Grant No. 11374005.

\end{document}